\documentclass{llncs}

\usepackage{llncsdoc}
\usepackage{amsmath}
\usepackage{cite}
\usepackage{amssymb}

\usepackage{graphicx}
\usepackage{graphics}

\usepackage{caption}
\usepackage{subcaption}
\usepackage{setspace}

\newcommand{\iseij}[6]{\left\{
    \begin{array}{ccc}
      {#1}&{#2}&{#3}\\
      {#4}&{#5}&{#6}
    \end{array}\right\}}

\title{The Screen representation of spin networks. Images of $6j$ symbols and semiclassical features}

\author{Mirco Ragni\inst{1}\and
Robert G. Littlejohn\inst{2}\and 
Ana Carla P. Bitencourt\inst{1}\and
Vincenzo Aquilanti\inst{3,4} \and
Roger W. Anderson\inst{5}}

\institute{Departamento de F\'isica, Universidade Estadual de Feira de Santana, Brazil \and
Department of Physics, University of California, Berkeley, California 94720, U.S.A \and
Dipartimento di Chimica, Universit\`a di Perugia, Italy, \email{vincenzoaquilanti@yahoo.it}\and
Istituto Metodologie Inorganiche e Plasmi CNR, Roma, Italy \and
Department of Chemistry, University of California, Santa Cruz, CA 95064, U.S.A., \email{anderso@ucsc.edu}
}

\begin{document}

\maketitle

\begin{abstract}
This article presents and discusses in detail the results of extensive exact calculations of the most basic ingredients of spin networks, the Racah coefficients ( or Wigner $6j$ symbols), exhibiting their salient features when considered as a function of two variables - a natural choice due to their origin as  elements of a square orthogonal matrix - and illustrated by use of a projection on a square "screen" introduced recently. On these  screens, shown are images which provide a systematic classification of features previously introduced to represent the caustic and ridge curves ( which delimit the boundaries between oscillatory and evanescent behaviour according to the asymptotic analysis of semiclassical approaches). Particular relevance is given to the surprising role of the intriguing symmetries discovered long ago by Regge and recently revisited; from  their use, together with other newly discovered properties and in conjunction with the traditional combinatorial ones, a picture emerges of the amplitudes and phases of these discrete wavefunctions, of interest in wide areas as building blocks of basic and applied quantum mechanics.
\end{abstract}

\section{Introduction}\label{sec:s1}
In this paper, extensive computational results serve to illustrate the main features of the well known Wigner $6j$ symbols ( or equivalently of the related Racah coefficients). Their importance has transcended the context of the quantum theory of angular momentum, where they were introduced originally: they appear as the building blocks of spin network structures, of widespread relevance in quantum science and its applications \cite{abfmr.08},\cite{abfmr.09}.

In their introduction as matrix elements between alternative angular momentum coupling schemes, Wigner and Racah had the insight of associating the six entries of a 6-j symbol with the lengths of the edges of a (generally irregular) tetrahedron and established asymptotic (or semi-classical) relationships with the geometrical properties, such as volumes and dihedral angles, of such a tetrahedron. In 1968, Ponzano and Regge \cite{ponzregge} initiated the study of the functional dependence of the $6j$-symbol on one of its six entries, arguing for a role of the tetrahedral volume and dihedral angles in the amplitude and phase of a (discrete) WKB-type of approximation of a wave-function. Independently, Neville \cite{nev} and Schulten and Gordon \cite{schgorb} provided rigorous derivations: the latter also introducing efficient computational procedures \cite{schgora} (for account of progress see \cite{ragni2010}) and numerical illustrations from this one-dimensional perspective (see also \cite{acc.01},\cite{littlejohn2009uniform},\cite{jpa2012}). The closely related Racah polynomials \cite{ac.00}, \cite{fcv.03}, \cite{avcf1995},\cite{askey} are at the foundation of modern approaches in the theory and applications of special functions and orthogonal polynomials. 
In this account, we present for the first time illustrations from the two-dimensional perspective, which is naturally based  on the view \cite{nev},\cite{littlejohn2009uniform},\cite{jpa2012} of the 6-j symbols as matrix elements enjoying a self dual property. The basic ideas of this approach are referred to in \cite{littlejohn2009uniform},\cite{jpa2012} as the 4-j model : accordingly here  plots as a function of two discrete variables are given in a square "screen" (see \cite{bitencourt2012exact}), in a sense generalizing the traditional presentations in square numerical tables \cite{varsh}. After a presentation of the general case in Section II, we illustrate symmetric and limiting cases in Section III (an important case being that of the Clebsch-Gordan coefficients, also known as Wigner's 3-j symbols, see \cite{aquilanti2007semiclassical} ). In the order of presentation, we are closely following the previous classification of the classical-quantum boundaries \cite{bitencourt2012exact}. In section IV, we provide additional and concluding remarks.

\section{Some theory and methods}
We present screen images in the next section that include the values of the $6j$ symbols or more precisely the $U(x,y)=\sqrt{\left(2x+1 \right)\left(2y+1 \right)}\iseij{a}{b}{x}{c}{d}{y}$.  The $U$ values have been calculated with a variety of methods:  direct summation with multi-precision arithmetic \cite{307}, \cite{3xx}, exact integer arithmetic, three- and five-term recursion relations \cite{lncs1.2013}, and checked by solving the eigenvalue equation \cite{lncs1.2013}.  All the calculations give precise agreement with each other.

\subsection{Canonical ordering}
We choose the Canonical ordering of the $a,b,c,d$ as proposed in \cite{lncs1.2013}.  In this ordering $a$ is the smallest of the eight values $a,b,c,d,a',b',c',d'$ with the primed quantities the Regge conjugate values of the unprimed values.  The ordering assures that the screen has dimension $(2a+1)\times(2a+1)$ and the ranges $b-a \leq x \leq b+a$ and $d-a \leq y \leq d+a$.  For most of the cases considered in this paper, this Canonical form agrees with a slightly different ordering proposed in \cite{aquilanti2013volume}.  

We have also found that the following expressions for $s, r, u$, and $v$ are useful for describing the topology of the screens corresponding to different values for $a, b, c$ and $d$.  See Ref. \cite{aquilanti2013volume} for more discussion.

\begin{eqnarray}\label{eq:uterms}
s &\equiv& \left[(a+c)+(b+d)\right]/2 \label{eq:s} \\
r &\equiv& \left[(a+c)-(b+d)\right]/2 \label{eq:r}  \\
u  &\equiv& \left[(a+b)-(c+d)\right]/2 \label{eq:u} \\ 
v  &\equiv& \left[(a+d)-(b+c)\right]/2 \label{eq:v}
\end{eqnarray}

For these definitions, $s$ is the semiperimeter, $r$ is the difference in the sums of column values, $u$ is the difference in the sums of row values, and $v$ is the difference between sums of diagonals.  The definitions for $r,u,v$ in equations \ref{eq:r}, \ref{eq:u}, and \ref{eq:v} constrain the values for $c$ in the Canonical ordering  such that either $v$ or $r$ must be equal or less than $0$, and $u$ must also be less than $0$.    

\subsubsection{Ponzano-Regge Theory}

The screen images show many features most of which are explained with the Ponzano-Regge theory and some symmetry considerations.  The Ponzano-Regge estimate for $6j$ in the classical region ($V^2 > 0$) is
\begin{equation} \label{eq:PRs}
\iseij{a}{b}{x}{c}{d}{y} \approx \frac{1}{\sqrt{12 \pi \vert V \vert}} \cos\left(\Phi\right),
\end{equation} .

Hence the $6j$ symbols have a magnitude envelope determined by the tetrahedron volume, $V$, and oscillations given by the Cosine of the Ponzano-Regge phase $\Phi$.  Both the volume and the phase are given by the geometry of the tetrahedron with sides: $A,B,C,D,X$, and $Y$ (See \cite{ponzregge} for details.  Here $A=a+1/2, ...,X=x+1/2,Y=y+1/2$).

The square volume of the tetrahedron can be calculated with a Cayley-Menger or a Gram determinant, but the results obtained with both determinants are 
equivalent to the famous formula known to Euler but first found
five centuries ago by the Renaissance mathematician, architect and
painter Piero della Francesca. We give his formula arranged as
needed in the following.

\begin{eqnarray}\label{eq:piero}
288V^2 & = & 2A^2C^2 (-A^2+B^2+X^2+Y^2+D^2-C^2) \nonumber \\
       &   & + 2B^2D^2 (A^2-B^2+X^2+Y^2-D^2+C^2)\nonumber\\
       &   & + 2X^2Y^2( A^2+B^2-X^2-Y^2+D^2+C^2) \nonumber \\
       &   & - (A^2+C^2)(B^2+D^2)(X^2+Y^2)\nonumber\\
       &   & - (A^2-C^2)(B^2-D^2)(X^2-Y^2)
\end{eqnarray}
We write the Ponzano-Regge phase as $\Phi = A\theta_1 + B\theta_2 + X\theta_3 +C\eta1 +D\eta_2+Y\eta_3 + \frac{\pi}{4}$ (See \cite{ponzregge}, \cite{schgorb}).

\subsection{Piero line symmetries}
An important screen symmetry is the possible presence of a Piero line where the $6j$ and $U$ are symmetric with respect to interchange of $x$ and $y$. 
The classical symmetry relation: $\iseij{a}{b}{x}{c}{d}{y} = \iseij{a}{d}{y}{c}{b}{x}$ shows that the screen will be invariant to this interchange if $b = d$, if the $6j$ is written in the conventional order where $a$ is equal to the smallest argument.  The Piero line is the diagonal corresponding to $x = y$, and the $6j$ or $U$ are symmetric with respect to this line.  The Piero equation for $V^2$, Eq. (\ref{eq:piero}) shows this symmetry very clearly.  The first four lines in Eq. (\ref{eq:piero}) are symmetric with respect to interchange of $X$ and $Y$, but the symmetry is broken with the term in the last row unless $B=D$.  (It seems that the other possible case where $A=C$, can not occur for conventional ordering unless $B$ is also equal to $D$).   There appear no exact symmetries for $U$ or $6j$ with respect to the line $x + y = x_{min} + y_{max}$. There will be a Piero line symmetry whenever $u = v$ (See Eqns. \ref{eq:u}, \ref{eq:v}).  Piero line symmetries are found in Figures \ref{fig.01d}, \ref{fig2r},  \ref{fig3r}, \ref{fig4r}, and \ref{fig6r} in Section \ref{sec:s8}.

\subsection{Regge symmetry}
The Regge conjugate will be the same as the original if the sum of any two of $a,b,c,d$ is the same as the sum of the other two.  Hence the Regge conjugate will be the same if  the product $ruv=0$. This equivalence between the original and Regge conjugate is found for the cases corresponding to Figures \ref{fig.01b}, \ref{fig.01c}, \ref{fig.01d}, \ref{fig3r}, \ref{fig4r}, \ref{fig5r}, and \ref{fig6r}.
 
\subsection{Location of Gates}

We define a gate as the place along each of the four sides of the screen where the caustic line, $V^2=0$ touches the side.  The gate will generally be some where in the center of a side, but for specific choices for $a,b,c,d$ the gates may coalesce at the corners of the screen or occupy an entire side. Eqns. \ref{eq:r}, \ref{eq:u}, and \ref{eq:v} also yield information about the location of the gates. We start with the approximate equation (\ref{eq:yvmax1}) for the value of $y$ that gives the maximum volume of the tetrahedron for a given $x$.  This approximate equation assumes that all of the quantities $A,B,C,D,X,Y$ are large enough to be replaced by $a,b,c,d,x,y$. 

 \begin{equation}\label{eq:yvmax1}
  y^2_{Vmax} = \frac{(a^2-b^2)(c^2-d^2)+
  (a^2+b^2+c^2+d^2)x^2-x^4}{2x^2}.
 \end{equation}
 
This equation gives the analytic results that the values of $y$ giving positive $V^2$ in the corners of the screen are given as follows (We are assuming conventional ordering).  These results were first described in Ref. \cite{bitencourt2012exact}.
\begin{enumerate}
\item
For $r=0$ (Eq. \ref{eq:r}), For $x=b-a,~y_{Vmax}=d-a$.  Positive $V^2$ is found at the south-west corner.  As $|r|$ increases, the lower branch of the caustic line is found further from this corner.  See figure \ref{fig.01d}.
\item
For $u=0$ (Eq. \ref{eq:u}), For $x=b+a,~y_{Vmax}=d-a$. Positive $V^2$  is found at the south-east corner.  See figure \ref{fig.01c}.
\item
For $v=0$ (Eq. \ref{eq:v}), For $x=b-a,~y_{Vmax}=d+a$. Positive $V^2$ is found at the north-west corner.  See figure \ref{fig.01b}.
\end{enumerate}
We can note that positive $V^2$ will never occur in the north-east corner.  
More than one of these conditions may be present.  See figures \ref{fig3r} and \ref{fig4r} for $u=v=0$,  figure \ref{fig5r} for $r = v = 0$ ( here the west side of the caustic line is the line $x=b-a$, and figure \ref{fig6r} for $r=u=v=0$.

\section{Images and Discussion}\label{sec:s8}

This section reports and discusses a series of graphs where the plots
of caustics and ridges published in Ref. \cite{bitencourt2012exact} are superimposed on $x-y$ color plots of true $U$.  The following features are common to all of the plots except Figures \ref{fig7}, \ref{fig_signed.01a}, and \ref{fig:ridges}. 
\begin{enumerate}
\item
The angular momenta $a,b,c,d$ are written in Canonical form, which means that the screens have dimension $(2a+1)\times(2a+1)$, and the ranges of $x$ and $y$ are $b-a \leq x \leq b+a$ and $d-a \leq y \leq d+a$.
\item
All the figures show the caustic line (light gray oval)that encircles the central, classical regions where ($V^2 > 0$).
\item
All the figures also show the ridge lines solid and dashed white lines.
\item
All of the figures except Figures \ref{fig7}, \ref{fig_signed.01a}, and \ref{fig:ridges} display $|U(x,y)|$.
\end{enumerate}

\begin{figure}
\centering
  \includegraphics[width=.85\textwidth]{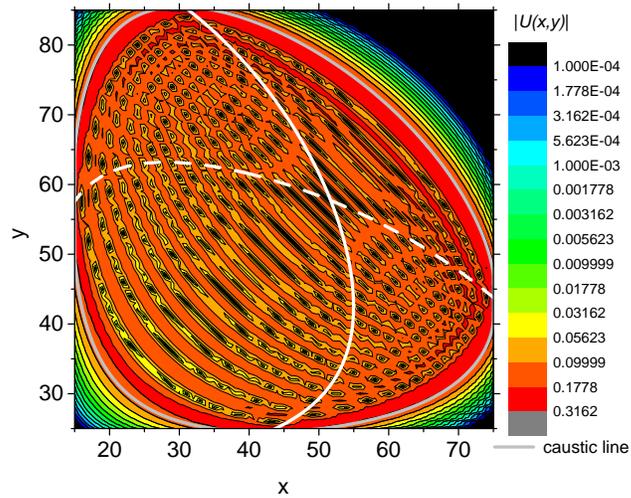}\\
\caption{Plot of $|U (x,y)|$ for $a = 30, b = 45, c = 60, d = 55$. 
Ranges are $15 \leq x \leq 75$ and $25 \leq y \leq 85$. 
There are $2a+1$ values of $x$ and $y$.  This corresponds to fig. 1a in \cite{bitencourt2012exact}. The caustic closed curve (light gray), corresponding to zero volume (Eq. \ref{eq:piero}) encircles the "classical" region of positive volume, showing oscillatory behaviour, while outside in the four "nonclassical" regions the values are exponentially decaying,. This is the "canonical" form with $a \leq b \leq d$ namely the screen is oriented, the caustic touching the sides at four points, denoted North, West, South and Eastern gates( see Sec Concluding remarks). The corresponding  Regge conjugate can be shown to be $a' = 50, b' = 65, c' = 40, d' = 35$. None of the primed quantities are smaller than $a$.
}\label{fig.01a}
\end{figure}

\subsection{General Case}
\begin{figure}
\centering
  \includegraphics[width=.85\textwidth]{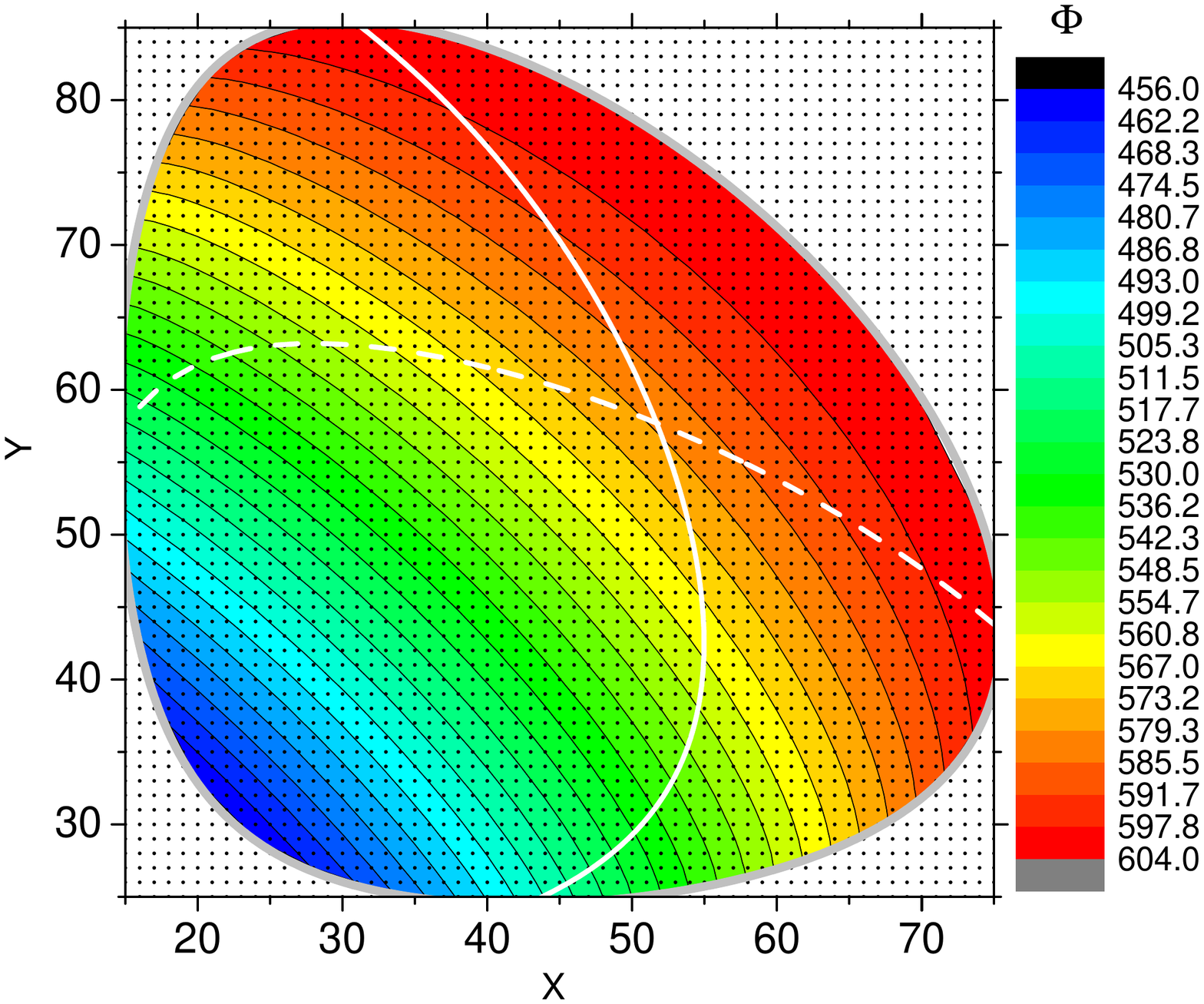}\\
  \caption{The figure shows the region of the $x,y$ plane
relevant for the $6j$-symbol with the same values of $a$, $b$,
$c$ and $d$ as in Fig. \ref{fig.01a}.  The small spots
are the quantized values of $x$ and $y$, at which the
$6j$-symbol is defined.  The heavy light gray oval curve is the caustic line,
which surrounds the classically allowed region. The lighter lines and color changes
inside the classically allowed region are the contours of the
Ponzano-Regge phase $\Phi$. Contour lines are separated by a phase difference of $2\pi$.  As we move across a horizontal line, varying $x$ while holding
$y$ fixed, we can see how many spots lie between two contour
values of $\Phi$.  For example, near the upper right side of the
caustic curve there are up to nine spots between contour values.
This means that if the $6j$-symbol is plotted in a stick diagram,
as in Fig. 1 of Ref. \cite{littlejohn2009uniform}, then there will be
several sticks under a single lobe of oscillation of the
$6j$-symbol, as shown in the right side of that figure.  But near
the bottom of the caustic curve, there are approximately only two
spots per $2\pi$ increment of phase, which means that the sticks
alternate in sign, as shown on the left side of Fig. 1 of
Ref. \cite{littlejohn2009uniform}.
}\label{fig7}
\end{figure}

\begin{figure}
\centering
  \includegraphics[width=.85\textwidth]{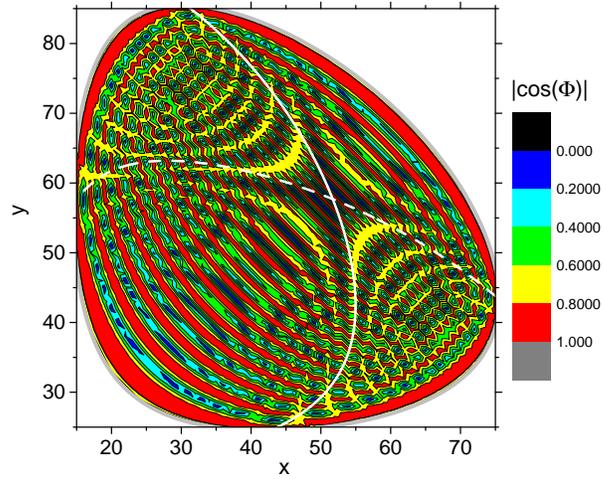}
\caption{$|\cos(\Phi)|$ for angular momenta of Figure \ref{fig.01a}.  This figure is based on the $\Phi$ in figure \ref{fig7}.}
\label{fig_signed.01a}
\end{figure}

\begin{figure}
\centering
  \includegraphics[width=.85\textwidth]{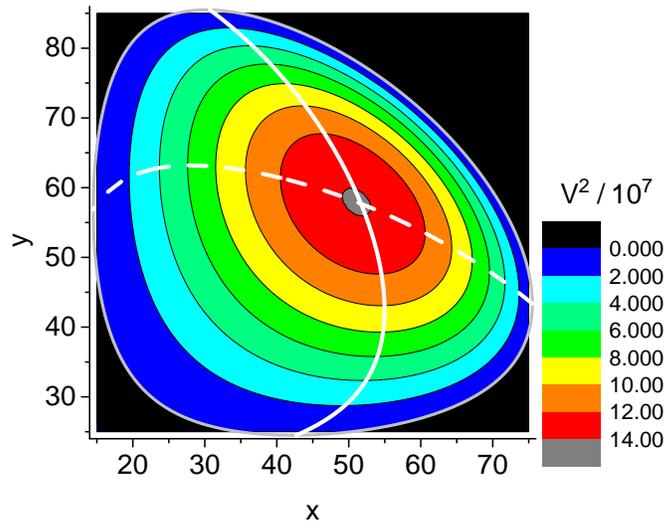}
\caption{Volume, caustics, ridges for angular momenta of Figure \ref{fig.01a}.} \label{fig:ridges}
\end{figure}

\begin{figure}
      \centering
      \begin{minipage}[b]{0.48\textwidth}
         \centering
         \includegraphics[width=\textwidth]{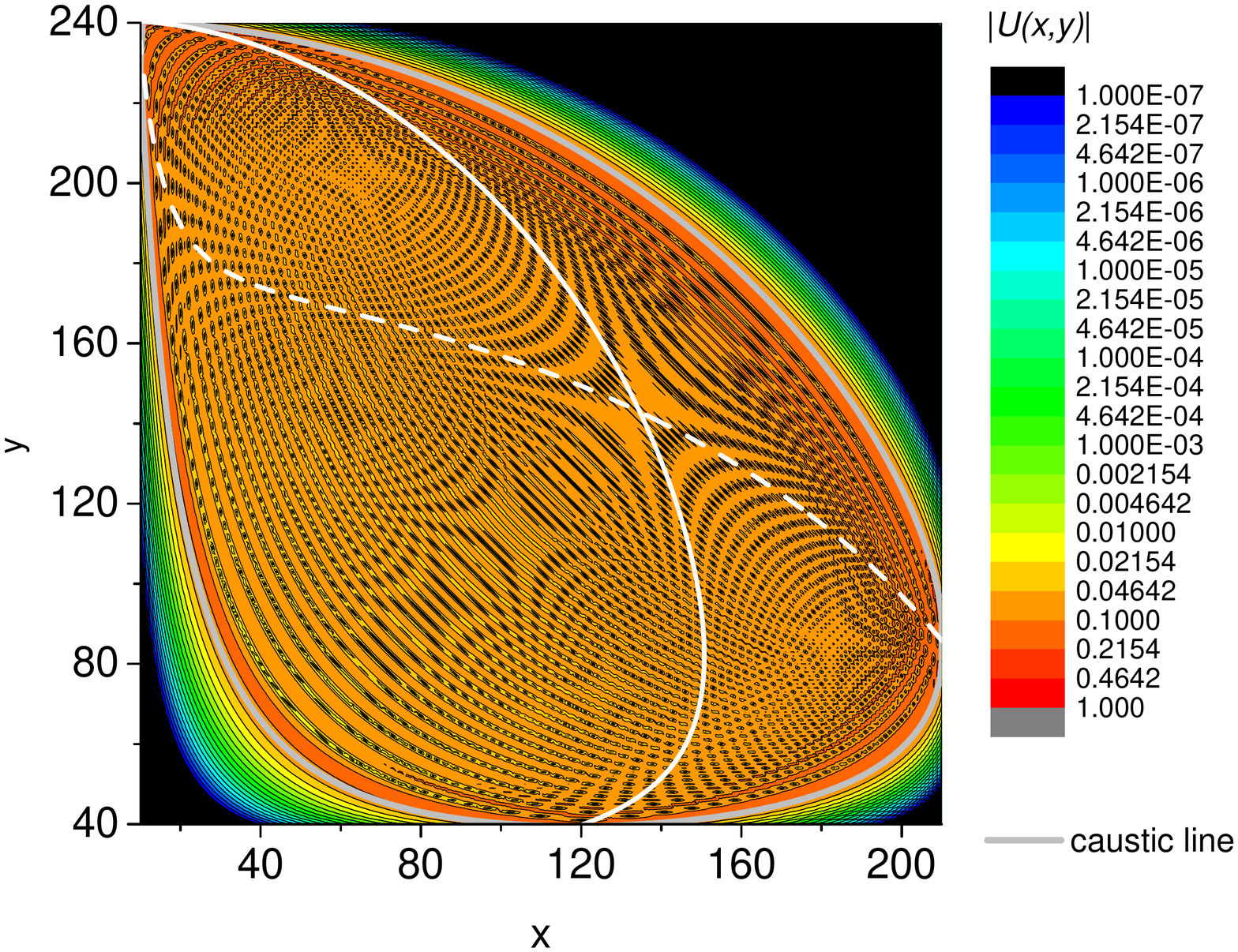}
        \subcaption{ Plot of $|U (x,y)|$ for $a = 100, b = 110, c = 130$ and $d = 140$. Ranges are $10 \leq x \leq 210$ and $40 \leq y \leq 240$.  This corresponds to fig 1b in \cite{bitencourt2012exact}. Caustic and ridge lines are shown. . In this case $v = 0$ and therefore the two Regge conjugates are identical. Note the coalescence of Northern and Western gates at the upper left corner, also because $v=0$, Eq. \ref{eq:v}.}\label{fig.01b} 
      \end{minipage}%
     ~~ \begin{minipage}[b]{0.48\textwidth}
        \centering
         \includegraphics[width=\textwidth]{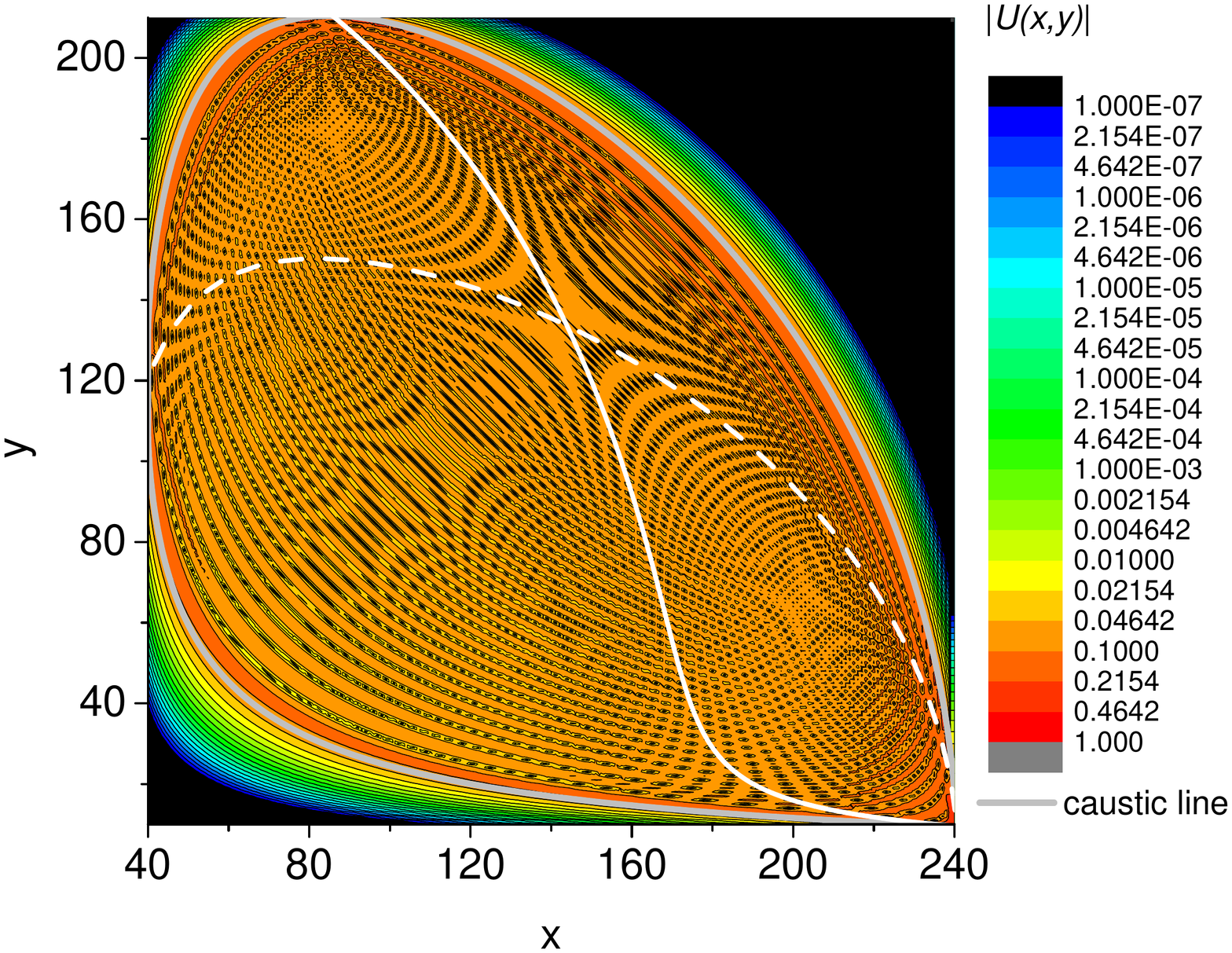}
         \subcaption{Plot of $|U(x,y)|$ for $a=100$, $b=140$, $c=130$ and $d=110$. $40 \leq x \leq 240$ and $10 \leq y \leq 210$.  This corresponds to fig. 1c in \cite{bitencourt2012exact}. 
As in the previous case Fig. \ref{fig.01b}, the values spanned by $x$ and $y$ are both $2a+1$, but they are interchanged, namely the convention for the orientation of the screen is not adopted. Again, since $u = 0$, the two Regge conjugates are identical. Actually the previous case and this one are connected by a classical exchange symmetry, and the two figures are related by reflection with respect to the diagonal of the screen connecting lower left and upper right corners. Now the coalescence is between the East and South gates, which are moved to the lower right corner of the screen, because $u=0$, Eq. \ref{eq:u}.}\label{fig.01c}
    \end{minipage}  
   \caption{}
\end{figure}

\begin{figure}
      \centering
      \begin{minipage}[b]{0.48\textwidth}
         \centering
         \includegraphics[width=\textwidth]{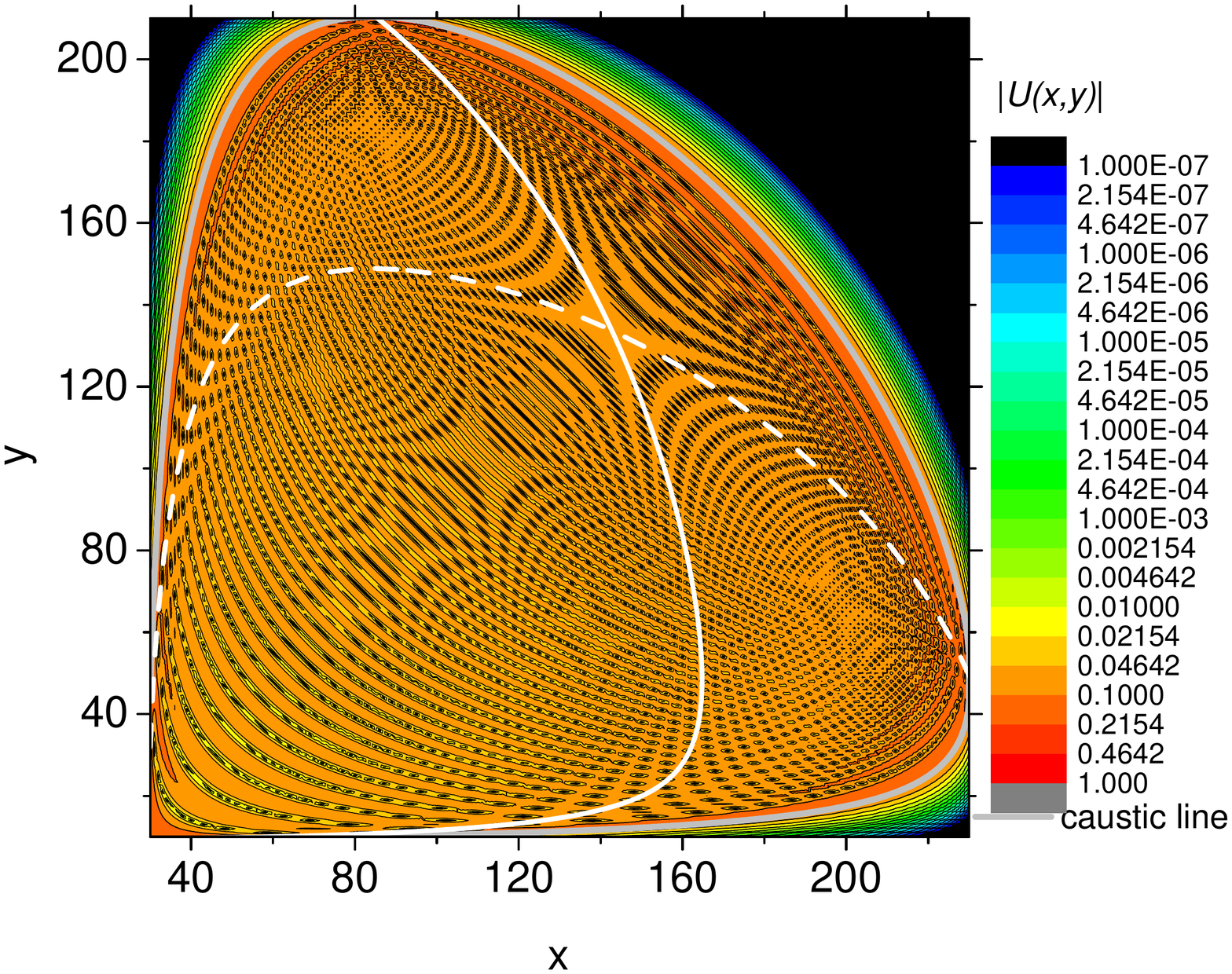}
         \subcaption{Plot of $|U(x,y)|$ for $a=100$, $b=130$, $c=140$ and $d=110$. $30 \leq x \leq 230$ and $10 \leq y \leq 210$. This corresponds to fig.1d in \cite{bitencourt2012exact}, where only the caustic and ridge curves were given. The values spanned by $x$ and $y$ are both $2a+1$. As in the previous two cases, a relationship holds: here we have $r = 0$ (Eq. \ref{eq:r}), and the two Regge conjugates are again identical, but  the coalescence is now between the West and South gates, in the lower left corner of the screen. A reflection symmetry is to be noted with respect to the diagonal of the screen connecting lower left and upper right corners. }\label{fig.01d}
      \end{minipage}%
     ~~ \begin{minipage}[b]{0.48\textwidth}
        \centering
         \includegraphics[width=\textwidth]{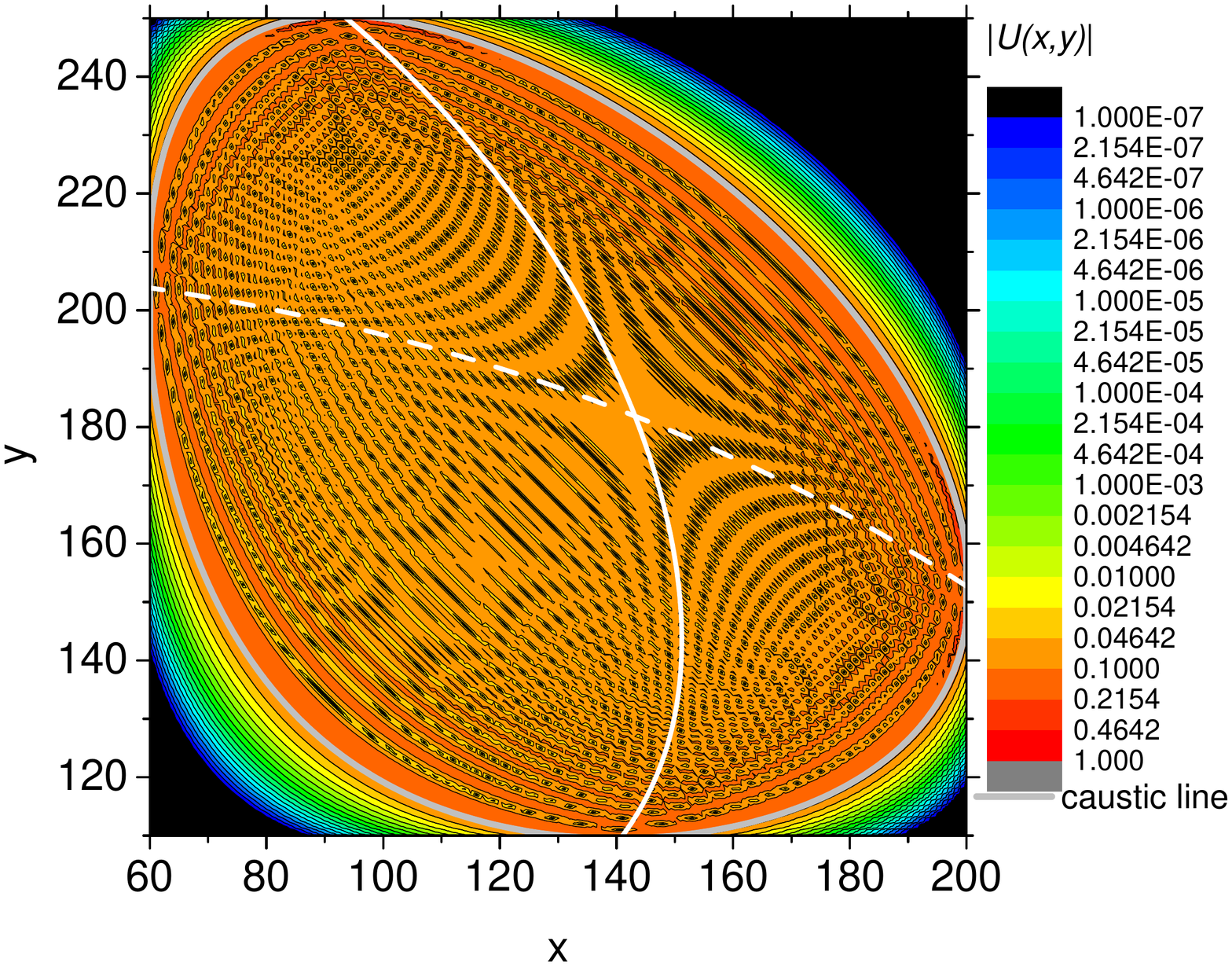}
           \subcaption{Plot of $|U(x,y)|$ for $a=70$, $b=130$, $c=180$ and $d=180$. $60 \leq x \leq 200$ and $110 \leq y \leq 250$. This plot of $|U (x,y)|$ corresponds to Fig. 2 in \cite{bitencourt2012exact}, where only the caustic and ridge curves were given for $a = 100, b = 100, c = 150$ and $d = 210$. Ranges are $60 = x = 200$ and $110 = y = 250$. This latter plot is not canonical: in fact, the Regge conjugate has $a' =70; b' = 130; c'= 180, d' = 180$. }
\label{fig1r}
    \end{minipage}  
   \caption{}
\end{figure}

\begin{figure}
\centering
  \includegraphics[width=\textwidth]{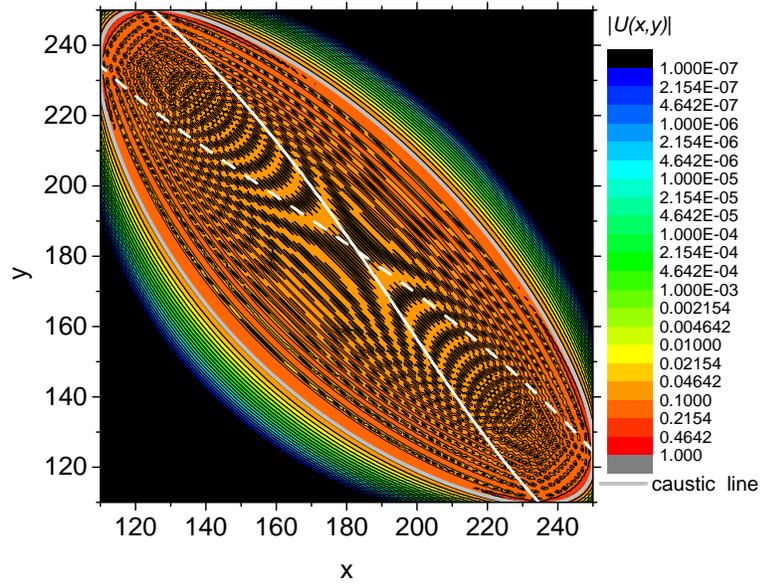}\\
  \caption{Plot of $|U(x,y)|$ for $a=70$, $b=180$, $c=130$ and $d=180$. $110 \leq x \leq 250$ and $110 \leq y \leq 250$. Plot of $|U (x,y)|$ corresponding to Fig. 3 in \cite{bitencourt2012exact}, where only the caustic and ridge curves were given, for $a = 100; b = 150; c = 100; d = 210$.  Since the Regge conjugate has $a'=70; b' = 180; c'= 130$ and $d' = 180$, the plot can be taken as "canonical", at variance with Fig 7. Now since $u=v=-30$  there is a Piero line which is the diagonal from the lower left corner to the upper right corner, making the plot symmetrical by reflection.}\label{fig2r}
\end{figure}

\begin{figure}
      \centering
      \begin{minipage}[b]{0.48\textwidth}
         \centering
         \includegraphics[width=\textwidth]{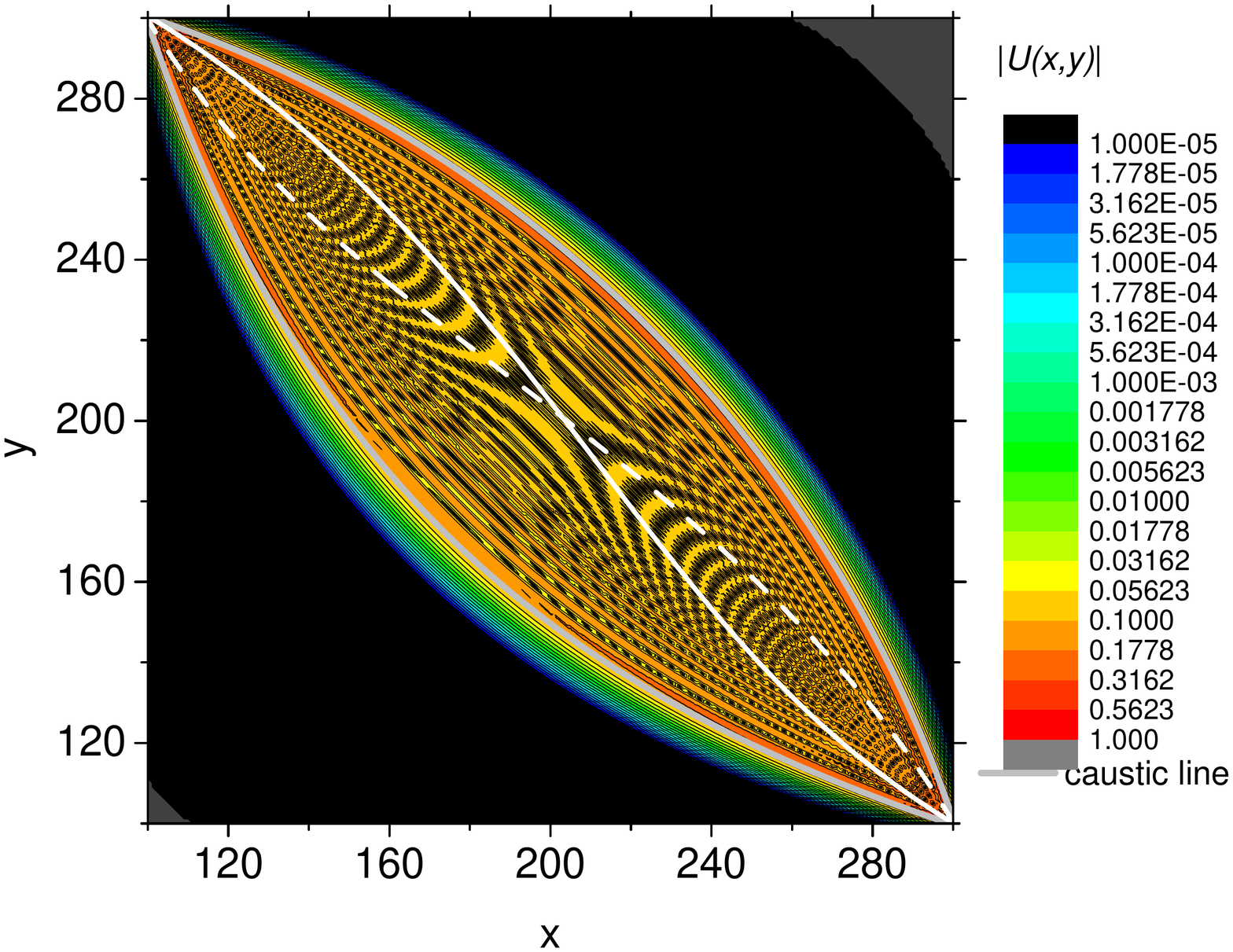}
      \subcaption{Plot of $|U(x,y)|$ for $a=100$, $b=200$, $c=100$ and $d=200$. $100 \leq x \leq 300$ and $100 \leq y \leq 300$.  Plot of $|U (x,y)|$ for the case of Fig. 4a in \cite{bitencourt2012exact}, where only the caustic and ridge curves were given, the canonical form being endorsed  when parameters are rewritten exchanging columns as follows: $a = 100; b = 200; c = 100$ and $d = 200$.  Here, since  $u=v=0$, Regge symmetry makes conjugates identical, and there is a Piero line symmetry.  Note for comparison to the following case that $r = -100$.  }\label{fig3r}
      \end{minipage}%
     ~~ \begin{minipage}[b]{0.48\textwidth}
        \centering
         \includegraphics[width=\textwidth]{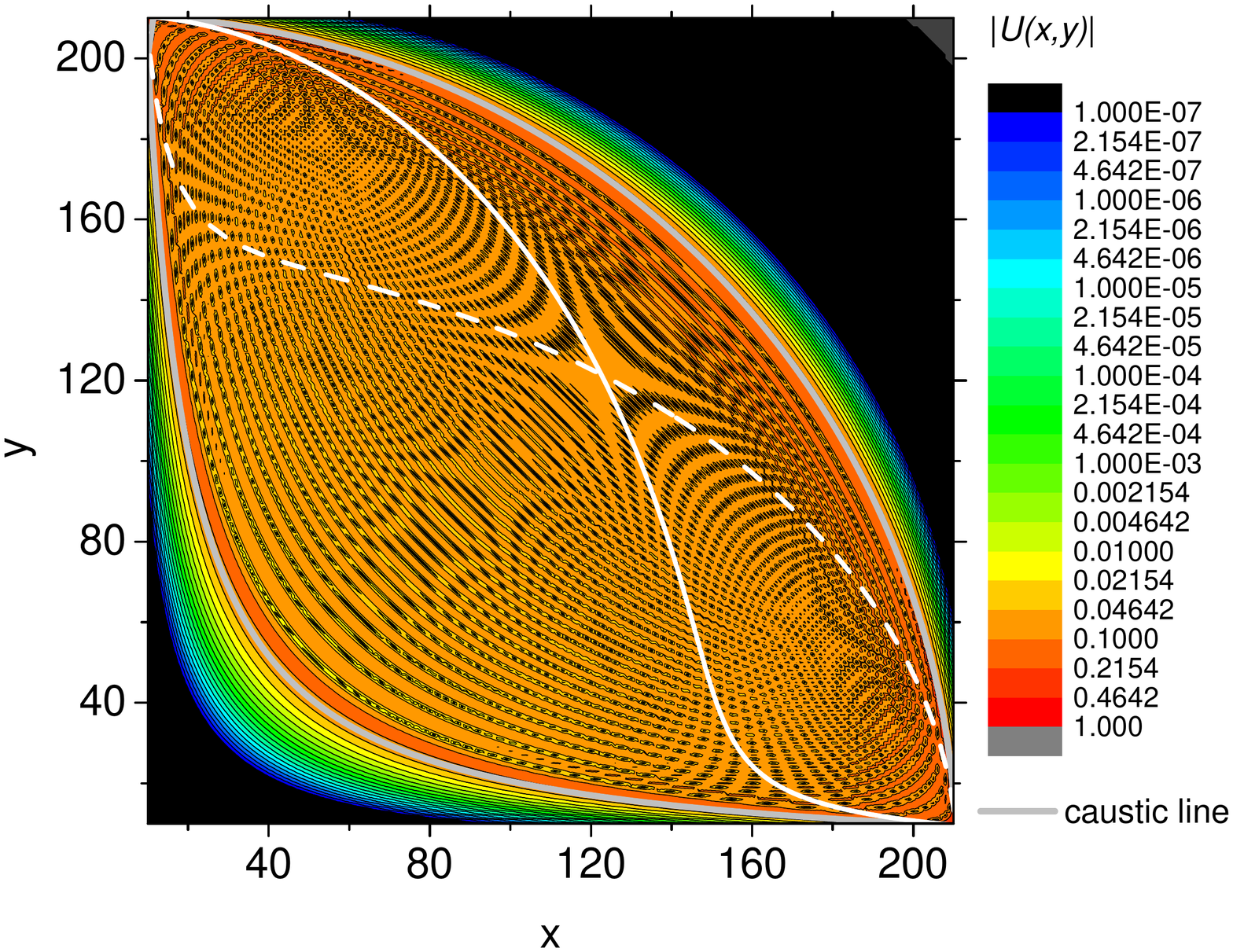}
          \subcaption{Plot of $|U(x,y)|$ for $a=100$, $b=110$, $c=100$ and $d=110$. $10 \leq x \leq 210$ and $10 \leq y \leq 210$.  Plot of $|U (x,y)|$ for the case of Fig. 4b in \cite{bitencourt2012exact}, where only the caustic and ridge curves were given, the canonical form being endorsed  when parameters are rewritten exchanging columns as follows: $a = 100; b = 110; c = 100$ and $d = 110$.  As in Fig 9, since $u = v = 0$,  Regge symmetry makes conjugates identical, and there is Piero symmetry.  Note for comparison to the previous case that now $r = -10$ :the lower difference between sums of columns shows qualitative shape changes for mismatch between columns.}\label{fig4r}     
    \end{minipage}  
   \caption{}
\end{figure}

\begin{figure}
      \centering
      \begin{minipage}[b]{0.48\textwidth}
         \centering
         \includegraphics[width=\textwidth]{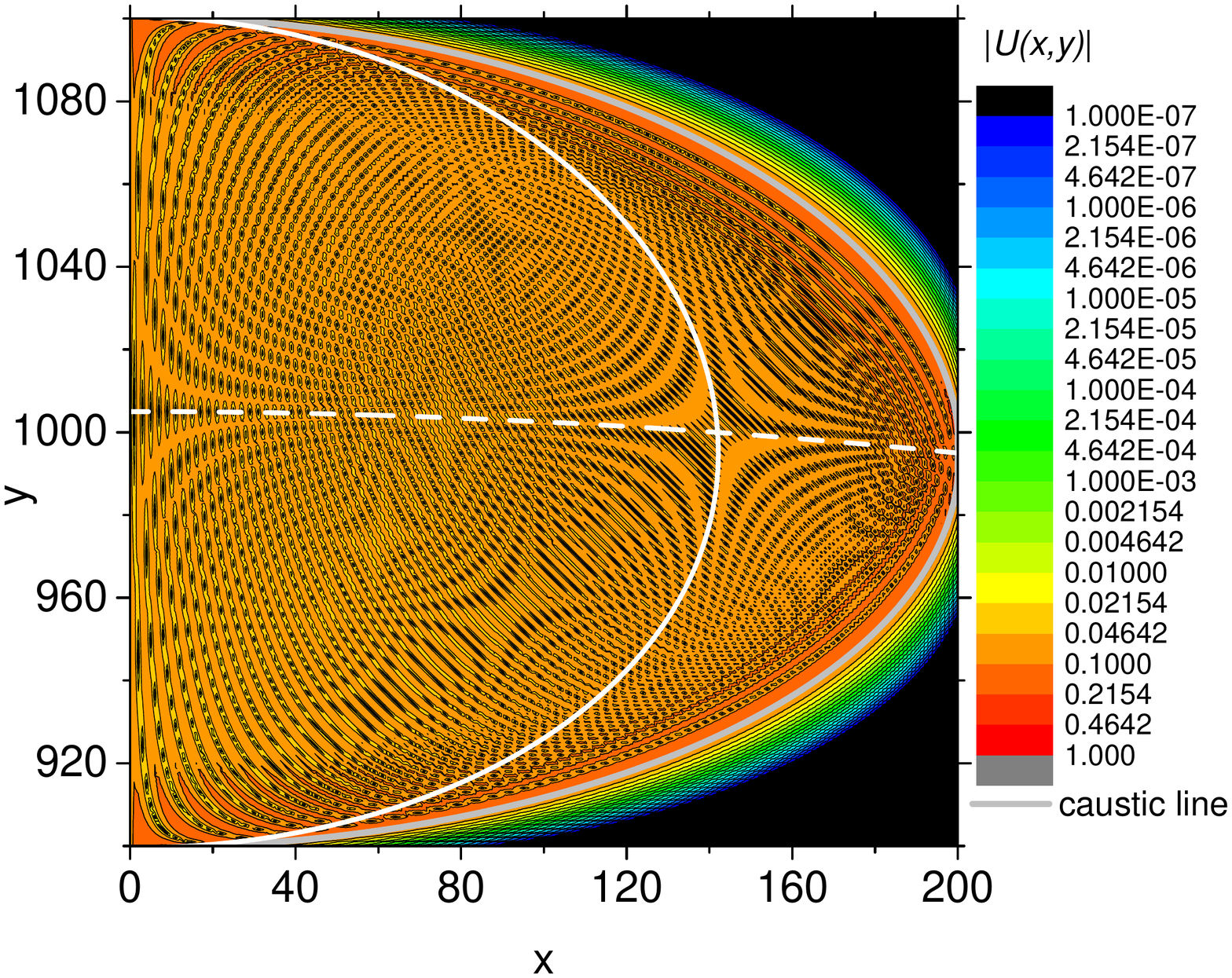}
           \subcaption{Plot of $|U(x,y)|$ for $a=100$, $b=100$, $c=1000$ and $d=1000$. $0 \leq x \leq 200$ and $900 \leq y \leq 1100$.
Case when both $r=0$ and $v=0$.  This corresponds to fig.5 in \cite{bitencourt2012exact}, where only the caustic and ridge curves were given.  Now $r=v=0$, and the two Regge conjugates are again identical, but  the coalescence is now of  both the North and South gates with the West gate, on the full line from the lower left to the upper left corners of the screen, As noted in \cite{bitencourt2012exact}, since $a, b$ and $x$ are smaller than $c, d$ and $y$, we can regard this plot as that of a $3j$ symbol,
( : : : ) where the entries in the upper row are the angular momenta $100, 100, x$ and the corresponding projections in the lower row are $y$ \textbf{–-} 1000, 1000 \textbf{--} $y$, 0. Note that a reflection along the $y$ line by mirror symmetry would lead to a replica of the image on the screen whereby the plane would consist of a classically allowed region limited by an ellipse  as a caustic curve. In the view of the plot as that of the $3j$ symbol, described above, the operation corresponds to that of allowing one of the nonzero projection to change sign.}\label{fig5r}
      \end{minipage}%
     ~~ \begin{minipage}[b]{0.48\textwidth}
        \centering
         \includegraphics[width=\textwidth]{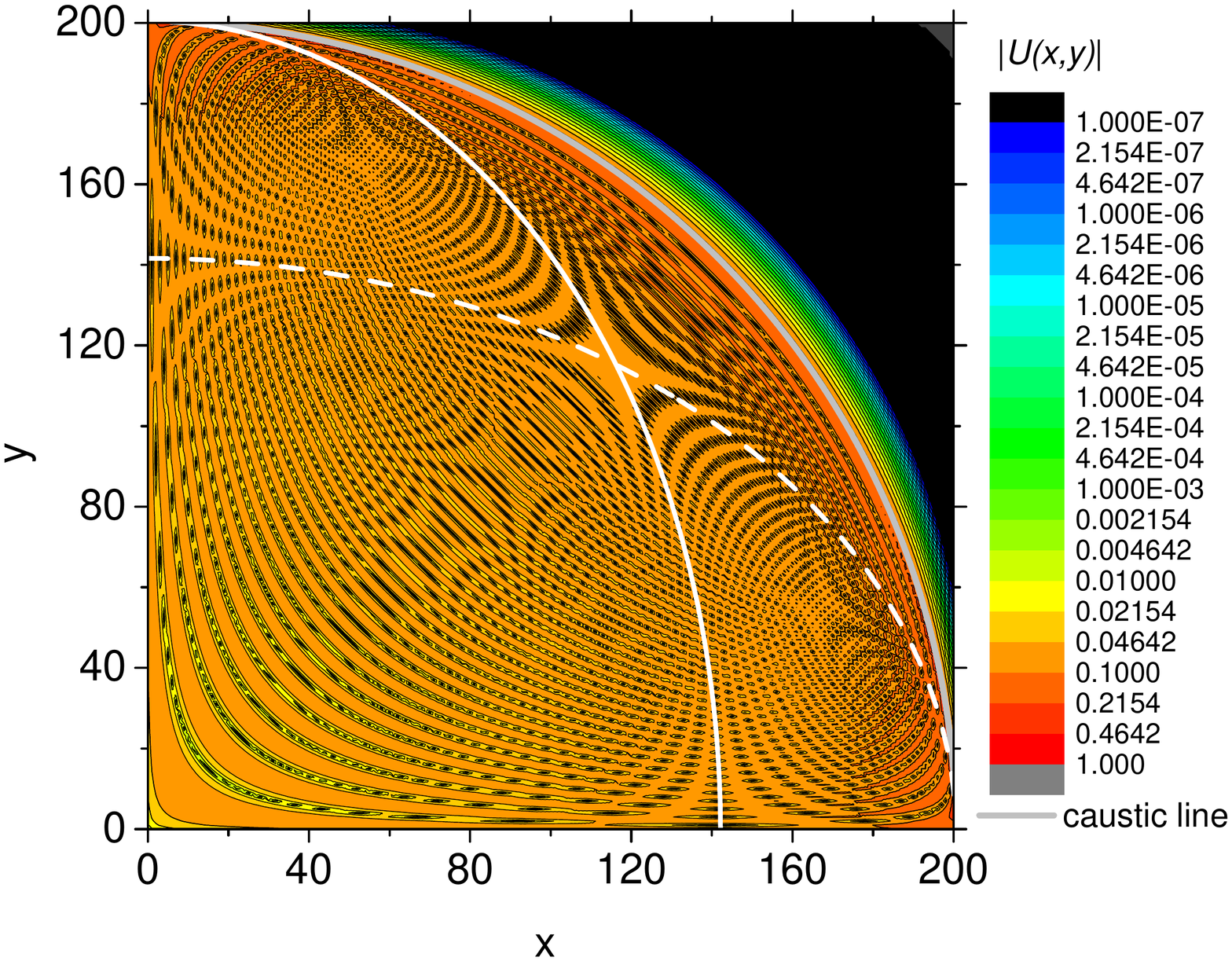}
         \subcaption{Plot of $|U(x,y)|$ for the fully symmetric case: $a=100$, $b=100$, $c=100$ and $d=100$. $0 \leq x \leq 200$ and $0 \leq y \leq 200$.  This corresponds to fig.6 in \cite{bitencourt2012exact}, where only the caustic and ridge curves were given.  The three relationships as in figures 4, 5 and 6 occur here, since $r = u = v = 0$. The two Regge conjugates are again identical: one of the coalescences is again as in Fig 11 of  both the North and South gates with the West gate, on the full line from the lower left to the upper left corners of the screen; :now another coalescence is of  both the East and West gates with the South gate, on the full line from the lower left to the lower right corners of the screen. Also since $ruv = 0$ there is a Piero line as the diagonal from the lower left corner to the upper right corner, making the plot symmetrical by reflection along this line. As noted in \cite{bitencourt2012exact}, repeated  reflections along the $x$ and $y$ lines by mirror symmetry would lead to replicas of the image on the screen, whereby the plane would consist of classically allowed regions limited by circles as caustics, tangent in four points.}\label{fig6r}
    \end{minipage}  
   \caption{}
\end{figure}

The general case is found when there are no special symmetries, and Figure \ref{fig.01a} shows an example.  This plot of $|U(x,y)|$ shows several striking features that can be explained inside the caustics line (boundary between $V^2 \geq 0$ and $V^2 < 0$) with the Ponzano-Regge theory \cite{ponzregge}. The figure shows that the $U$ are small where $V^2$ is maximum.  This occurs at the intersection of the two ridge lines, which show the  This occurs where the $V^2$ is maximum as a function of $y$ for given $x$ and as a function of $x$ for given $y$ (see Figure \ref{fig:ridges}).  The magnitude of $U$ tends to increase as the caustic line is approached the $x,y$ point corresponding to the maximum.  Most of the structure in Fig. \ref{fig.01a} is a consequence of the Ponzano-Regge phase (Cosine term in equation \ref{eq:PRs}).  Figure \ref{fig7} shows this phase for the same values of $a,b,c,d$.  Figure \ref{fig_signed.01a} gives the $|\cos\Phi|$ values for the screen, and it is obvious that the structure in Fig. \ref{fig.01a} agrees well with the expectations of the Ponzano-Regge theory.  We have found that the same argument can also explain all of the images in Figures 5-9 in this paper.

Figure \ref{fig.01a} also shows the general fact that the magnitude of the $U$ are oscillatory in the classical region, $V^2 \geq 0$, and exponentially decreasing as $x,y$ is moved deeper into the nonclassical region.  The values for $U$ can be estimated with a suitable extension of the Ponzano-Regge theory (See Ref. \cite{schgorb}.

\subsection{Symmetric cases}
Figures 5-9 show images of screens with different symmetries.  They illustrate cases where the gates  coalesce in the northwest, southwest, and southeast corners, and Piero line symmetries.  They also show cases where the Regge conjugates are the same as the $6j$ with the original arguments, and where the $6j$ approximate $3j$ symbols.

\section{Discussion, additional and concluding remarks}\label{sec:s7}
The extensive images of the exactly calculated $6j$'s on the square
screens illustrate how the caustic curves separate the classical
and nonclassical regions, where they show wavelike and evanescent
behaviour respectively. Limiting cases, and in particular those
referring to  $3j$ and Wigner's $d$ matrix elements can be analogously
depicted and discussed. Interesting also are the ridge lines,
which separate the images in the screen tending to qualitatively
different foldings of the quadrilateral, namely convex in
the upper right region,  concave in the upper left and lower right
ones, and crossed in the lower left region.

\textit{Catastrophe theory classification.}  The pictures of 6-j on the screen in the previous Section exhibit most clearly features amenable to be classified in terms of catastrophe theory, with a panorama of valley bottoms, ridges and both elliptic and hyperbolic umbilics arising in the two-dimensional membrane-like modes. See \cite{littlejohn2009uniform}, \cite{gilmore}.

\textit{Chirality gates.}  This remark concerns the formal analogy between the present problem of four angular momenta arranged as vectors having a (not necessarily) planar quadrilater structure and those of the motion of tetra-atomic or four center structures where bonds can be treated as \textit{rigid} while bending and torsion modes are allowed. If we consider $A,B,C,D$, as the lengths of the four bonds, and $X$ and $Y$ as the diagonals of the quadrilateral, i.e. the distances between atoms not connected by bonds, a \textit{screen} representation can be set up, the \textit{caustic} corresponding to allowed planar configurations. It is known that transition between chirality pairs in tetrahedral structure corresponds to \textit{flattened} structures, and therefore the \textit{caustic} curves shows configurations through which such a system would find its way to chirality exchange modes. In particular the four points where in the generic case (figures 1-4) the caustic touches the screen, are labeled accordingly North, West, South and East \textit{gates}, since they mark where and how a planar structure should fold to perform such chirality interchange mode. In Ref. \cite{aquilanti2013volume} a similarity is also pointed out with the celebrated problem of the kinematics of the four-bar linkage, the fundamental mobile mechanism of engines.

\textit{Alternative mappings.}  Motivated from the phase-space analysis of semiclassical dynamics in Ref. \cite{jpa2012}, alternatively to the $x,y$ \textit{screen} it is interesting to consider other conjugated variables such as e.g. $x$ and the associated momentum $p_x$, corresponding to a dihedral torsional mode. The corresponding mapping, rather than on a square, is on a spherical triangle on the surface of the sphere $S^2$. We are also exploring a third  type of mappings involving the modes of torsion angles corresponding e.g. to $p_x$ and $p_y$, of interest for intramolecular dynamics.

Work on extensions to $3nj$ symbols, to $q$ analogues, and to alternative coordinates of elliptic type is in progress.

\bibliographystyle{splncs}

\bibliography{aquila,sort}

\end{document}